\documentclass[conference]{IEEEtran}
\IEEEoverridecommandlockouts
\usepackage{cite}
\usepackage{amsmath,amssymb,amsfonts}
\usepackage{graphicx}
\usepackage{textcomp}
\usepackage{xcolor}
\def\BibTeX{{\rm B\kern-.05em{\sc i\kern-.025em b}\kern-.08em
    T\kern-.1667em\lower.7ex\hbox{E}\kern-.125emX}}
\usepackage{multirow}    					
\usepackage{hhline}     				 		
\usepackage{algorithm} 
\usepackage{algpseudocode}
\usepackage{enumerate}
\usepackage{wrapfig}
\usepackage{fancybox}
\usepackage[font={small}]{caption}
\usepackage{url}
\usepackage{balance}
\usepackage{amssymb,amsmath}
\usepackage{pifont}
\usepackage{xcolor}
\usepackage{tikz,pgfplots}
\usepackage{graphicx}
\usepackage{color}    
\usepackage{listings}
\usepackage{soul} 
\usepackage{subfigure}
\usepackage{booktabs}
\usepackage{array}
\usepackage{amssymb}
\usepackage{longtable}
\usepackage{framed}
\usepackage[most]{tcolorbox}
\begin{document}

\title{Dataset Bias in Android Malware Detection\\
}

\author{\IEEEauthorblockN{1\textsuperscript{st} Yan Lin}
\IEEEauthorblockA{
\textit{Beijing University of Posts and Telecommunications}\\
linyan@bupt.edu.cn}
\and
\IEEEauthorblockN{2\textsuperscript{nd} Tianming Liu}
\IEEEauthorblockA{
\textit{Monash University}\\
Tianming.Liu@monash.edu}
\and
\IEEEauthorblockN{3\textsuperscript{rd} Wei Liu}
\IEEEauthorblockA{\textit{Tsinghua University} \\
lw21@mails.tsinghua.edu.cn}
\and
\IEEEauthorblockN{4\textsuperscript{th} Zhigaoyuan Wang}
\IEEEauthorblockA{\textit{Beijing University of Posts and Telecommunications} \\
wangzgy@bupt.edu.cnD}
\and
\IEEEauthorblockN{5\textsuperscript{th} Li Li}
\IEEEauthorblockA{\textit{Monash University} \\
li.li@monash.edu}
\and
\IEEEauthorblockN{6\textsuperscript{th}Guoai Xu}
\IEEEauthorblockA{\textit{Beijing University of Posts and Telecommunications} \\
xga@bupt.edu.cn}
\and
\IEEEauthorblockN{6\textsuperscript{th}Haoyu Wang}
\IEEEauthorblockA{\textit{Huazhong University of Science and Technology} \\
haoyuwang@hust.edu.cn}
}

\maketitle
\thispagestyle{plain}
\pagestyle{plain}

\begin{abstract}
Researchers have proposed kinds of malware detection methods to solve the explosive mobile security threats. We argue that the experiment results are inflated due to the research bias introduced by the variability of malware dataset. We explore the impact of bias in Android malware detection in three aspects, the method used to flag the ground truth, the distribution of malware families in the dataset, and the methods to use the dataset. 
 We implement a set of experiments of different VT thresholds and find that the methods used to flag the malware data affect the malware detection performance directly. 
 We further compare the impact of malware family types and composition on malware detection in detail. The superiority of each approach is different under various combinations of malware families. Through our extensive experiments, we showed that the methods to use the dataset can have a misleading impact on evaluation, and the performance difference can be up to over 40\%.
 We argue that these research biases observed in this paper should be carefully controlled/eliminated to enforce a fair comparison of malware detection techniques.
 Providing reasonable and explainable results is better than only reporting a high detection accuracy with vague dataset and experimental settings. 
\end{abstract}

\begin{IEEEkeywords}
Malware detection, Dataset bias, Mobile App, Android.
\end{IEEEkeywords}

\section{Introduction}

With the explosion of smartphones and mobile apps, the number of mobile malware has been growing rapidly as well~\cite{wang2018beyond}. Millions of malicious apps were identified every year, with complex malicious payload, and sophisticated evasion techniques.
The official Android app market, Google Play, was reported to be affected by malware from time to time~\cite{NewMalware, Joker}, which causes great risks to mobile users.

The increasing mobile security threats have attracted a large number of research efforts in our community. Researchers have proposed different kinds of malware detection techniques, e.g., signature-based approaches~\cite{signature1, signature2, signature3, signature4}, behavior-based approaches~\cite{behavior1, behavior2, behavior3, mariconti2016mamadroid}, and machine-learning based approaches~\cite{ml3, drebin, ml5, aafer2013droidapiminer}, etc. 
Almost all the existing studies have reported promising results. For example, Drebin~\cite{drebin} was reported to achieve a detection rate of 94\% in Android malware detection, ICCDetector~\cite{xu2016iccdetector} was reported to achieve an accuracy of 97.4\%, and MADAM~\cite{dini2012madam} reports to achieve an overall malware detection accuracy of 100\%.

\textit{Is Android malware detection a solved problem?}
Considering the high accuracy reported in previous studies, it seems that there exists very little room for improvement. 
However, in this paper, we argue that the effectiveness of malware detection approaches is highly related to the dataset used in the evaluation, and the experiment results are inflated due to research bias introduced by the variability of the malware dataset and the methods to use the dataset.

\textbf{Representative malware benchmarks.}
There are some widely used malware datasets in the research community. 
For example, MalGenome dataset~\cite{genome} was constructed in 2012, and it contains over 1,000 malware samples, which were used by hundreds of studies to evaluate their effectiveness on malware detection. Drebin~\cite{drebin} is another widely used benchmark, with over 5,000 malware samples. 
AMD~\cite{amd} is one of the largest malware datasets used in the research community, with over 24 thousand samples covering 71 malware families in total. RmvDroid~\cite{wang2019rmvdroid} takes advantage of the app maintenance behaviors of Google Play to help label malware, which contains 9,133 malware samples across 56 malware families. Besides, a number of papers~\cite{fan2018android, pendlebury2019tesseract, suarez2018eight} have created their own datasets based on existing anti-virus engines for malware analysis, detection, and classification.

\textbf{Research Bias introduced by dataset.}
Existing malware datasets were created by different research groups, and they were constructed using ad-hoc approaches followed by different criteria (e.g., labeling methods, family distribution, and splitting methods of training/testing), which may cause potential research bias\footnote{We use the term 'bias' to refer to, different criteria to \textit{create} and \textit{use} the malware dataset can have a misleading impact (bias) on evaluations.}.

\begin{table*}[t]
\newcommand{\tabincell}[2]{\begin{tabular}{@{}#1@{}}#2\end{tabular}}
\centering
\caption{A list of widely used Android malware benchmarks. The \emph{method/source} indicates the method used to flag the ground truth of each dataset. The \emph{\# Families} indicates the number of malware families in each dataset, either reported by the original author or classified by AVClass~\cite{2016AVclass}. The \emph{top-3 families} and \emph{top-3 types} indicate the most popular malware families and types in each dataset. The column \emph{adware} indicates whether each dataset treats adware as a kind of malware.}
\label{table:existingdataset}

\begin{tabular}{l|l|l|l|c|c|l|l}
\toprule

Dataset      & Time      & \# Malware & Method/Source   & \# Families & \# Top-3 Families & \# Top-3 Types & Adware \\ \midrule

MalGenome~\cite{genome}    & 2010-2012 & 1,234      & Security Reports  & 49 &  \tabincell{c}{ DroidKungfu (38\%)\\Basebridge (25\%) \\Geinimi (5\%)}     &  \tabincell{c}{Trojan (94\%)\\Exploit (3\%)\\Spyware (1\%)}     & NO  \\ \hline

Drebin~\cite{drebin}       & 2013      & 5,560      & VT (2 of 10 engines)  & 179 & \tabincell{c}{ FakeInstaller (17\%)\\DroidKungfu (12\%) \\Plankton (11\%)} & \tabincell{c}{Trojan (76\%)\\Malware (18\%)\\Exploit (2\%)} & NO   \\  \hline


Piggybacking~\cite{piggybacking} & 2016      & 1,136      & VT (\textgreater{}=1 engine)   &  100 & \tabincell{c}{Dowgin (24\%)\\Kuguo (22\%)\\Gingermaster (6\%)} & \tabincell{c}{Adware (64\%)\\Trojan (25\%)\\Spyware (2\%)}   & \checkmark
  \\ \hline

AMD~\cite{amd}    & 2010-2016 & 24,553     & VT (\textgreater{}=28 engines) &   71 &  \tabincell{c}{Droidkungfu (20\%)\\Airpush (8\%)\\Ginmaster (8\%)} &  
\tabincell{c}{Trojan (42\%)\\
Adware (34\%)\\
Exploit (11\%)
} & \checkmark \ \\ \hline

RmvDroid~\cite{wang2019rmvdroid} & 2014-2018 & 9,133 & \tabincell{l}{VT (\textgreater{}=10 engines)\\ \& Removed by Google Play} & 56 &  \tabincell{c}{Airpush (32\%)\\Mecor (11\%)\\Plankton (9\%)} & \tabincell{c}{Adware (79\%)\\Trojan (13\%)\\Spyware (5\%)} & \checkmark
 \\
\bottomrule
\end{tabular}

\end{table*}

First, \textbf{the methods used to flag the ground truth are ad-hoc and usually vary greatly}. Besides MalGenome~\cite{genome}, which was manually created by analyzing the security reports released by security companies, almost all the other malware datasets were created based on the detection results of VirusTotal~\footnote{https://www.virustotal.com}, a widely-used online malware detection service containing over 60 anti-virus engines. The general method is to collect mobile apps in the wild (e.g., Google Play and third-party app markets), and then use the detection results of VirusTotal to label them. However, there are no standards on how to take advantage of the detection results to label malware. In this context, researchers use their intuition and adopt ad-hoc methods to label the apps and release the dataset to the research community as benchmarks. As there are over 60 engines that report the detection results, they usually use different thresholds of detection engines on VirusTotal to label malware samples. For example, Drebin~\cite{drebin} was created based on the results of 10 engines on VirusTotal, i.e., one sample is selected as long as two of the 10 engines flagged the sample as malicious. 
The Piggybacking~\cite{piggybacking} dataset used 1 engine as the threshold, 
the TESSERACT~\cite{pendlebury2019tesseract} sets the threshold as 4,
and AMD~\cite{amd} used 28 engines (over 50\% of the engines) as the threshold. 
\emph{It is unknown to us to what extent do the malware labelling methods affect the malware detection results}.


Second, \textbf{the malware families distributed in the datasets are unbalanced and vary greatly}. 
For example, MalGenome~\cite{genome} dataset covers 49 families, and each family contains 1 to 309 malware samples. The top-3 families occupy roughly 70\% of the overall dataset, while over 30 families have less than 10 samples.
The distribution suggested that, as long as the detection approach can successfully detect the top families, the overall result will be good enough.
As machine-learning based malware detection approaches highly rely on the training dataset, it is unknown to us \emph{how does the construction of the dataset affect the malware detection results}.

Third, \textbf{the methods to use the dataset may affect the detection results.} Traditionally, machine learning-based approaches are assessed in a cross-validation scenario that validates the classification model by assessing how the result will generalize to an independent dataset. To estimate how the prediction model will perform in practice, a cross-validation scenario partitions the sample data into 2 subsets. The first subset is used for learning analysis, i.e., building the model during the training phase. The second subset is used to validate the model. 
To reduce the variability of the results, multiple rounds are performed and the results are averaged over the rounds. 
A well-known type of cross-validation is the 10-Fold cross-validation, i.e., the sample data was randomly partitioned into 10 subsets, 9 of which are used for training and 1 for validation.
However, no previous work has analyzed the bias introduced by ``randomly'' partitions. For example, there are 10 malware families in a malware dataset.
When performing malware detection (note here, we mention malware detection, not classification), the ideal case should be that, for each kind of malware, we have known its 90\% of samples, and predict the remaining 10\%. However, the extreme case might be that, training use samples of 9 families and the remaining 1 family is used for prediction.
Almost all previous studies have not carefully explained how they use the dataset, thus \textit{we argue that there might be research bias when evaluating different approaches}.

As aforementioned, almost all the existing malware detection studies directly use the labelled malware datasets in an ad-hoc way, without considering the research bias and potential impact introduced by the malware dataset.

In this paper, we take a data-driven method to research the research bias introduced by the variability of malware dataset. We mainly focus on three aspects, the method used to flag the ground truth, the distribution of malware families in the dataset, and the methods to use the dataset. 
We first resort to the Androzoo dataset to obtain a malware corpus containing 690,544 samples. Then we create a number of experimental datasets under different conditions to investigate the potential bias introduced by malware datasets.

First, we explore the impact of VirusTotal threshold on malware detection. We constructed 23 sets of malware benchmarks including the VT numbers from 1 to 30+. We further apply \texttt{csbd}, \texttt{drebin}, and \texttt{mamadroid} to these benchmarks.  Interestingly, we observe that the VT threshold could affect the malware detection directly. The malware detection results of the algorithm increase with the increasing of VT thresholds at first, and then stabilize to some extent. The evaluation of different algorithms may also differ when labeling malware with different VT thresholds.  
 
 Second, we compare the impact of malware family numbers, types and composition on malware detection in detail. We randomly select 10 popular malware families from the malware corpus. We then curate a number of malware datasets with different numbers or different compositions of malware families. We find that the performance of the malware detection method is very related to the constructed dataset. The malware detection results of an algorithm can vary when different datasets are constructed.
 
 Third, we explore the methods to use the dataset when evaluating the malware detection methods. To be able to present the results more visually, we consider the ideal case and the extreme case of the sample partition methods in cross-validation. We compare the ideal case with the extreme case using 5-Fold cross validation
 Through our extensive experiments, we showed that the methods to use the dataset can have a misleading impact on evaluation, and the performance difference can be up to over 40\%.

Our research work's contributions are summarized as the followings:
\begin{itemize}
    \item We make the first systematic study of the dataset bias in Android malware detection. Specifically, we focus on the variability of the malware dataset and the methods to use the dataset.
    \item We make effort to curate a number of malware benchmarks under different conditions and evaluate their performance gap of malware detection using existing techniques. We found that the detection results can even be manipulated by controlling the dataset to make one specific approach achieve the best result. 
    \item We measure the influence of inconsistent datasets on malware detection assessments. We believe that a machine learning fairness framework for malware detection is needed to highlight pros and cons of each approach, and provide reasonable and explainable results. 
\end{itemize}

This paper is organized as follows. In Section \ref{Related Work and Background}, we present the related work and background on Android malware benchmarks and Android Malware Detection. In Section \ref{sec:design}, we present our research questions and the experimental setup in detail. We analyze the impact of malware labelling methods in Section \ref{sec:rq1}. We then further characterize how the variation of malware families affects the malware detection (seeSection \ref{subsection:familyDistribution}). In Section \ref{sec:rq4}, we investigate the methods to use the dataset in the malware detection, focusing on the "randomly" partitions of the cross validation. In Section \ref{Discussion}, we discuss the implication and limitation of our paper. Finally, we draw our conclusion in Section \ref{Conclusion}.
\section{Related Work and Background}
\label{Related Work and Background}

\subsection{Android Malware Benchmarks}
\label{subsection:malwaredataset}

Our community has created a number of datasets to facilitate Android malware research. We have summarized some of them in Table~\ref{table:existingdataset}.

MalGenome dataset was created in 2012 based on manually examining the security reports released by anti-virus companies. It has over 1,200 malware samples, covering 49 families. 
Over 94\% of the samples in the dataset are Trojan, and adware is not considered to be malware in this dataset. 
The top-2 families, i.e., DroidKungFu and Basebridge occupy over 60\% of the samples. MalGenome was widely used by our research community as the benchmark to perform malware detection~\cite{shen2018android, yerima2018droidfusion, kim2018multimodal}.
Drebin is another widely used benchmark, which was created in 2013. It has over 5,000 malware samples, and it was created based on the detection results of 10 famous anti-virus engines on VirusTotal. One sample is considered to be malicious as long as two of the 10 engines flagged the sample as malicious.
Over 76\% of the samples are Trojan, and adware is also not considered in this dataset.
Piggybacking dataset was created based on the app clone detection results (i.e., malicious apps were created on top of popular legitimate apps) in 2016. Apps will be regarded as malicious as long as they are flagged by at least one engine on VirusTotal. It contains 1,136 samples, and adware occupies a large portion of the samples (64\%).
The AMD dataset was created in 2016, with a large number of malware samples. It contains a considerable number of samples overlapped with MalGenome and Drebin, as it collected samples from multiple sources including existing malware datasets.
RmvDroid was created in 2019, in order to overcome the challenge of malware sample labelling. They rely on both VirusTotal and Google Play’s app maintenance practice to create the dataset. For the apps flagged by over 20 engines on VirusTotal and further be removed from Google, they will regard them as malware. RmvDroid also considers adware as malicious, which occupies 79\% of the samples. 

Besides, there are some other datasets created by the research community using the similar approach, including Android PRAGuard dataset \cite{maiorca2015stealth}, AAGM dataset \cite{lashkari2017towards}, the Android Malware Dataset\cite{wei2017deep}, and KronoDroid dataset \cite{guerra2021kronodroid} etc. 

AAGM dataset \cite{lashkari2017towards} was constructed in 2017. Applications in AAGM dataset were labelled as malware as they were flagged by more than two Anti-Virus products in Virustotal web service.It has 1900 applications including benign and 12 different families. It contains 400 malware samples, adware(250) and general  malware(150). 
Android PRAGuard dataset \cite{maiorca2015stealth} contains 10479 samples. It has over 50 malware families. They did not use VirusTotal or AndroTotal because of their limitations for their study.
The Android Malware Dataset \cite{wei2017deep}, is a public dataset consisting of 24,553 malware samples. Each app in the Android Malware Dataset is scanned by 55 antivirus products from VirusTotal. The sample is labelled as malware when over 28 anti-virus products used in the VirusTotal recognize it as a malware. It has 71 malware families and collected between 2010 and 2016.
KronoDroid dataset \cite{guerra2021kronodroid} is a hybrid-featured Android dataset. The malware dataset of KronoDroid is composed of Drebin, AMD, VirusTotal Academic Malware Samples and VirusShare. Thus, all malware samples are labeled by the result of Virustotal with different Virustotal threasholds. It has two public datasets. The emulator dataset is composed of 28,745 malware from 209 malware families and the real device dataset contains 41,382 malware, belonging to 240 malware families.

As aforementioned, these datasets were created under different criteria, including malware labelling methods, dataset construction, etc. We argue that these different criteria may introduce research bias in malware detection, which is the main focus of this paper.In order to analyze the bias of datasets with different criteria, we did not use these public datasets directly. We reconstructed many datasets according to these criteria metioned above for the study in this paper.

\subsection{Android Malware Detection}
\label{subsection:related}

Many research efforts were focused on malware detection in our community. They mainly could be classified as signature-based approaches ~\cite{signature1, signature2, signature3, signature4}, behavior-based approaches~\cite{behavior1, behavior2, behavior3, mariconti2016mamadroid}, and machine-learning based approaches~\cite{ml3, drebin, ml5, aafer2013droidapiminer}, etc.

DroidAPIMiner \cite{aafer2013droidapiminer} performs malware detection based on features generated at API level. Drebin is a lightweight detection method that uses static analysis to gather the most important characteristics of Android applications including permissions, API calls, components, etc. 
It uses SVM algorithm to detect whether a given sample is malicious or not. 
Csbd~\cite{csbdpaper} was proposed to detect malware using structural features, i.e., CFG signatures. It constructs CFGs of individual methods and encodes them as text-signatures. Then, a Random Forest classifier is trained with these signatures to detect malware.
DroidSIFT \cite{signature3}  builds contextual API dependency graphs that provide an abstracted view of the possible behaviors of malware and employs machine learning and graph similarity to detect malicious applications. 
MudFlow \cite{avdiienko2015mining} leverages the analysis of flows between APIs to detect malware.
DroidDetector \cite{yuan2016droiddetector} build Deep Belief Networks for Android malware detection using human engineered features, including required permissions, sensitive API calls, and certain dynamic behaviors. Deep4maldroid \cite{hou2016deep4maldroid} constructs weighted directed graphs from Linux kernel system calls and use them to train deep neural networks for malware detection. Mclaughlin et al. \cite{ml5} proposed to detect malware by applying a deep neural network to the raw op-codes extracted from the dex bytecode.
Mamadroid \cite{mariconti2016mamadroid} is an Android malware detection system based on modeling the sequences of API calls as Markov chains. This detection system depends on app behavior and builds a model in the form of a Markov chain.

Although previous studies have reported promising results on malware detection, it is still unknown to us how the results vary across datasets constructed under different criteria.
In this paper, we select three representative works, \texttt{csbd}, \texttt{drebin}, and \texttt{mamadroid} to revisit their malware detection performance across different datasets we constructed.
\section{Study Design}
\label{sec:design}


\subsection{Research Questions}

In this paper, to unravel the confusion mentioned in \textbf{Section \ref{Related Work and Background}}, we seek to answer the following research questions:

\begin{itemize}
    \item[RQ1]  \textit{To what extent does VirusTotal threshold in malware labelling affect the malware evaluation results?}
    VirusTotal has been widely used in the research community to label Android malware. As there are over 60 anti-virus engines on VirusTotal, it is hard to define an accurate threshold to label malware. As previous work has applied different ad-hoc thresholds, it is interesting to investigate Whether such thresholds impact malware evaluation results.
    
    \item[RQ2] \textit{How does the malware family distribution in the dataset affect the detection result?} As shown in Table \ref{table:existingdataset}, the malware family distribution in the widely used dataset is highly \textit{imbalanced}. Thus, it is unknown to us whether the imbalance would introduce research bias during the evaluation of malware detection approaches.
    
    
    \item[RQ3] \textit{How does the sample partition methods in evaluation affect the detection results?} No previous work has analyzed the bias introduced by the so-called ``randomly'' partitions in the evaluation (e.g., cross-validation) scenario. We want to explore the impact introduced by partition methods used in malware detection. 
    
\end{itemize}

\begin{figure}[h]
	\centering
	\includegraphics[width=0.95\linewidth]{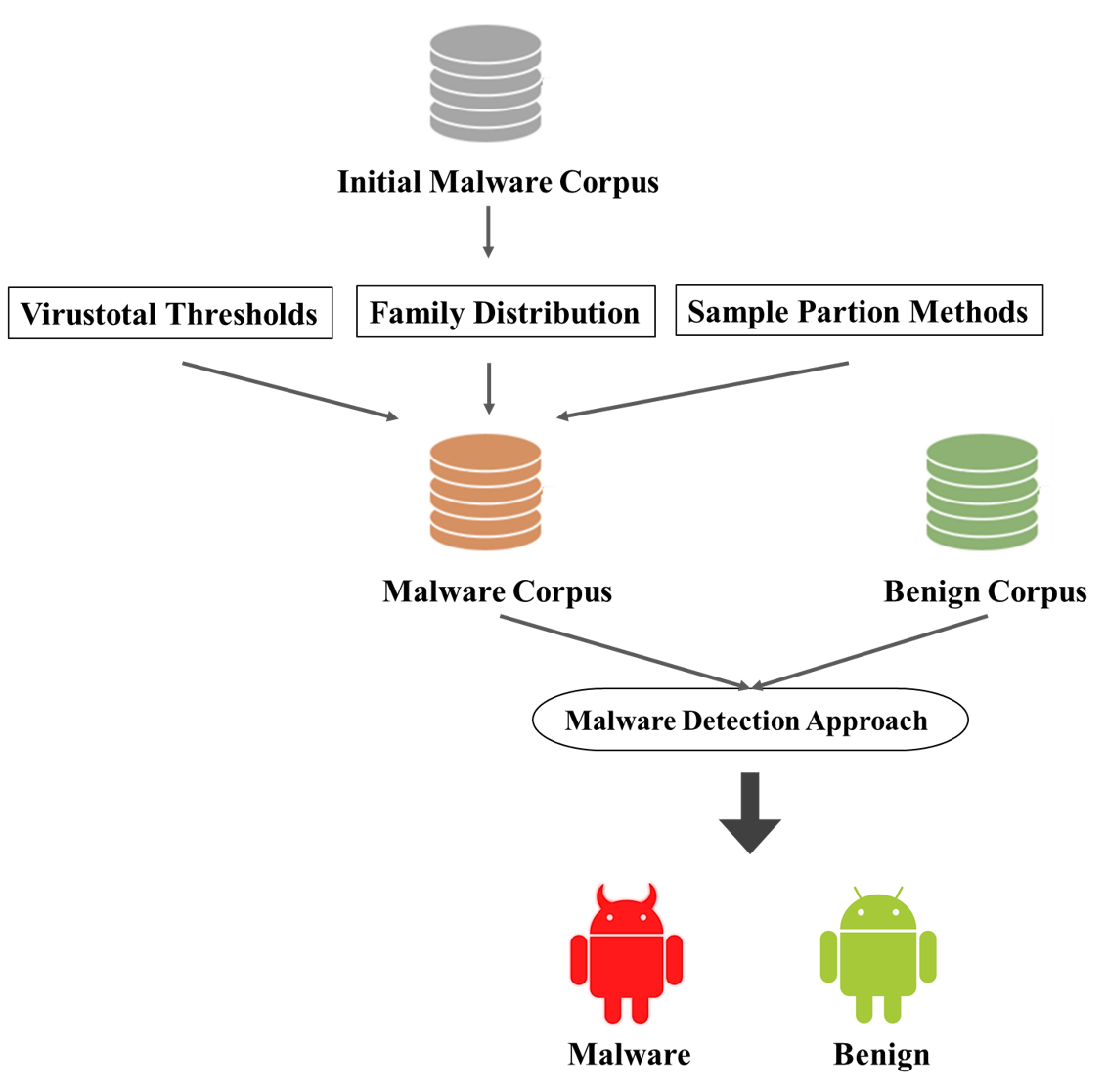}
	\caption{The overview of our experimental setup.}
	\label{fig:overview}
\end{figure}

\subsection{Experimental Setup}
\label{subsection:setup}

As shown in Fig. \ref{fig:overview}, we prepare the malware corpus based on Virustotal thresholds, family distribution, and sample partion methods. In this section, we illustrate our experimental setup in detail.

\subsubsection{Malware Detection Approaches}
To investigate the three RQs, in our experiment, we apply three widely used machine-learning based malware detection approaches (see Section \ref{subsection:related}), \texttt{csbd}, \texttt{drebin}, and \texttt{mamadroid}, as they are open-sourced, widely used in our community, and proven to be effective.
We reused the implementation from github for \texttt{csbd}\footnote{https://github.com/MLDroid/csbd}, \texttt{drebin}\footnote{https://github.com/MLDroid/drebin} and \texttt{mamadroid}\footnote{https://bitbucket.org/gianluca\_students/mamadroid\_code}, respectively.

\subsubsection{Malware Corpus}
To prepare our dataset for malware detection, we resort to the well-known Androzoo dataset~\cite{li2017androzoo++}, which contains over ten million Android apks from different app markets including Google Play. Androzoo provides the scanned result of each apk using VirusTotal.
According to these VT results, we are able to obtain 876,235 potential malicious apps that are originated from Google Play with at least one VT engine flagging them as positive by the time of this study. As previous work~\cite{label-dynamic} suggested that the detection results of VirusTotal would change with time, thus we further obtain their latest VT reports.
After fetching the latest VT results, the number of malware apks is reduced to 690,544, which means that almost 200K malware apks are not flagged as positive by any VT engines in the latest VT results. It further suggests the dynamic of VT and the potential inaccurate results in malware detection introduced by the unsteadiness of the dataset.

Based on these latest VT reports, we further obtain the family type and family name of each (potential) malicious app with Euphony~\cite{hurier2017euphony}. 
The 690,544 malicious apps, together with the VT number, family name, and family type of each malware, constitute the potential malware corpus in our experiment. Note that, apps in this malware corpus are not necessary to be malicious. 
We did not use all the samples as our ground-truth to evaluate the malware detection techniques. 
Based on the different criteria (e.g., VT threshold) and factors (e.g., malware family distribution) we considered, we will select apps from the corpus to construct different testing datasets.

\subsubsection{Testing Datasets}
To investigate the potential bias introduced by malware datasets, we have curated a number of experimental datasets under different conditions.
We first prepare a benign dataset of 1,000 apps, with no VT engines flagging them as positive in Androzoo all the time. 
Note that the benign dataset is used repeatedly throughout our experiments. 
For the malware dataset, we select 1,000 apps from our malware corpus based on different criteria (the selection process will be introduced in the following sections), and then combine them with the 1,000 apps in the benign dataset, forming a new experiment set of 2,000 apps. 
We then perform our experiments on these apps with a fully random 5-Fold cross-validation.
Note that in each iteration, our three malware detection approaches are applied respectively. Therefore, after one group of the experiment is conducted, we can obtain the accuracy, precision, and recall for the three approaches.
The only variable in each group of our experiment is how we select and construct the 1,000 malicious apps. We will elaborate on that in the following sections.
\begin{figure}[h]
\centering
\subfigure[Overall accuracy with different VT thresholds.]
{	\label{fig:rq1}
	\includegraphics[width=0.93\linewidth]{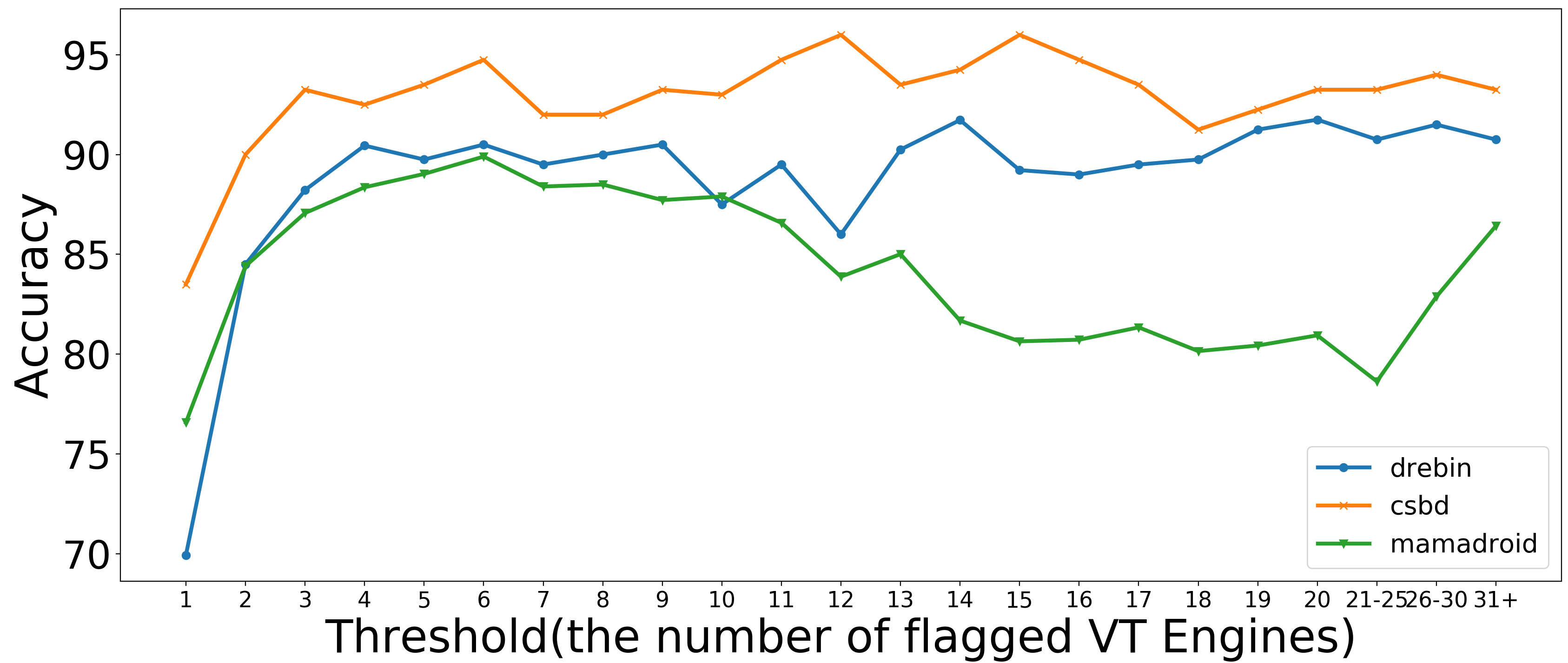}

}
\subfigure[Precision with different VT thresholds.]
{
	\label{fig:rq1-precision}
	\includegraphics[width=0.93\linewidth]{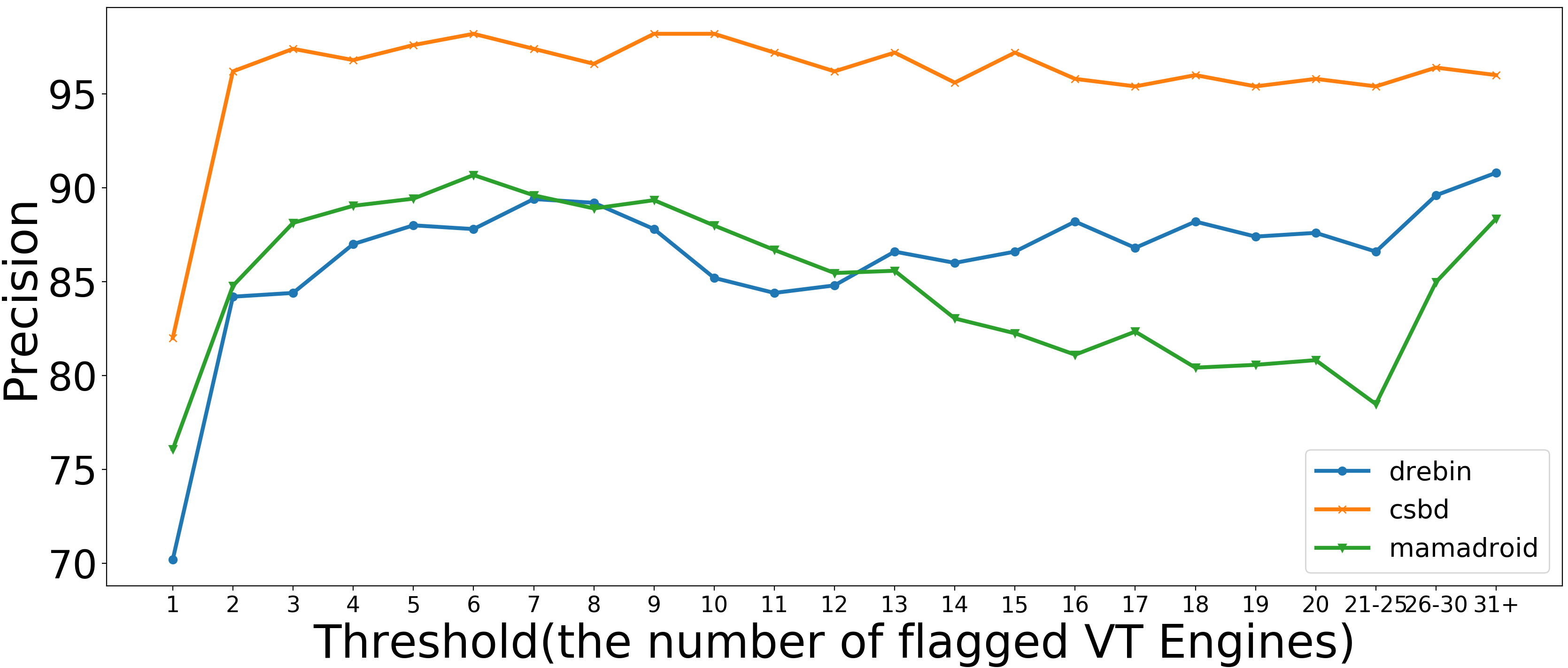}
}
\subfigure[Recall with different VT thresholds.]
{
	\label{fig:rq1-recall}
	\includegraphics[width=0.93\linewidth]{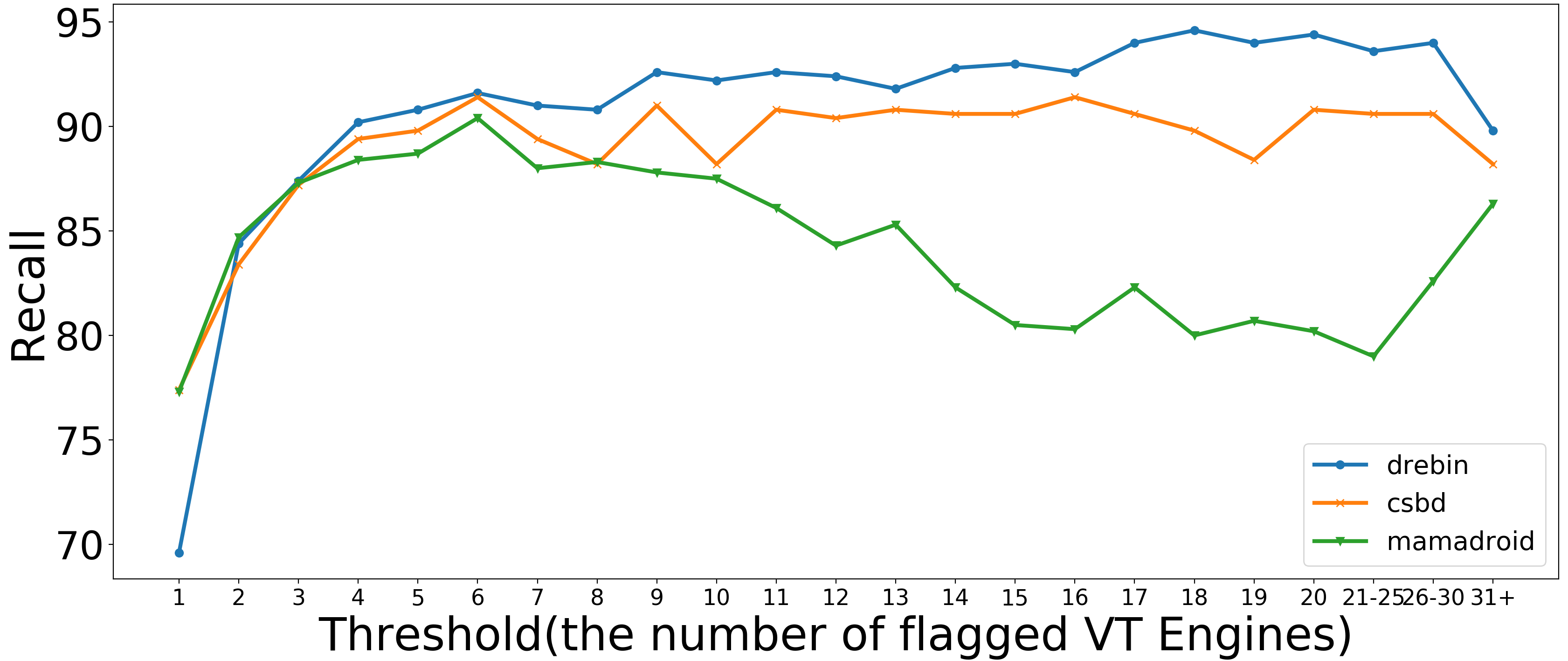}
}

\caption{Overall accuracy, precision, and recall of malware prediction for each group of experiments in RQ1.}
\label{fig:rq1all}
\end{figure}

\section{The Impact of Malware Labelling}
\label{sec:rq1}

\subsection{Experimental Design}
We first investigate the impact of VT threshold on malware detection. 
To provide a comprehensive study, we have constructed 23 sets of malware benchmarks, and each set contains 1,000 (considered) malicious apps based on different VT threshold.
Note that, when the VT threshold is over 20, the number of malware samples in our malware corpus is less than 1,000 for a single VT number. To ensure the same size of malware dataset, we extract the ones labeled as malware by 21-25 VT engines as a dataset. It is the same when the VT number is 26-30 and more than 31. The VT numbers of the 23 groups dataset range from 1 to 20 in different groups, with three additional groups of \emph{from 21 to 25}, \emph{from 26 to 30} and \emph{more than 31}. 
As aforementioned, we further apply \texttt{csbd}, \texttt{drebin}, and \texttt{mamadroid} to these benchmarks. 

\subsection{Analysis}
\textbf{Intra-approach Analysis}
As shown in Figure \ref{fig:rq1}, for each approach, the malware detection accuracy grows logarithmicly with the VT threshold increases from 1 to 6, and then the accuracy of \texttt{csbd} and \texttt{drebin} remains relatively stable from 6 to 31+, the accuracy of \texttt{mamadroid} remains relatively stable form 6 to 12. The difference between the highest and the lowest accuracy reaches up to 11.25\% for \texttt{csbd}, 21.5\% for \texttt{drebin}, and 13.33\% for \texttt{mamadroid} under different VT thresholds, which indicates that the criteria to construct the dataset would lead to the great diversity of the detection results.  

This is reasonable since a higher VT number represents an app having a higher possibility of being malicious, while a lower VT number may jeopardize the reliability of the ground truth in the experiment dataset, which in turn causes the decrease of the overall accuracy.

\textbf{Inter-approach Analysis}
In general, \texttt{csbd} has the highest accuracy, followed by \texttt{drebin} and then \texttt{mamadroid}. 
However, it is interesting to see that, the comparison of the three malware detection approaches on some metrics can show some inconsistent results introduced by the VT threshold.
For example, as to the malware detection precision, the performance of \texttt{mamadroid} is greater than that of \texttt{drebin} when the threshold is lower than 12, while the result shows a quick turnaround with a threshold higher than 12 (see Figure~\ref{fig:rq1-precision}). 
Further, the performance gaps among these approaches are highly sensitive to the threshold. For example, on the dataset of $VT=21-25$, \texttt{drebin} achieves 12.12\% of accuracy higher than \texttt{mamadroid} (90.75\% VS. 78.63\%), while they achieve similar results on the dataset of $VT=10$.

\begin{tcolorbox}[title= \textbf{Brief Summary}, left=2pt, right=2pt, top=2pt,bottom=2pt]
The thresholds for malware labelling can directly affect the performance of malware detection. For a single approach, the impact could be significant since the difference between the worst case and the best case reach over 20\%. Further, the malware labelling threshold can lead to \textit{controversial} or inconsistent results when comparing different malware detection approaches. 
\end{tcolorbox}





\section{The Impact of Dataset Composition}
\label{subsection:familyDistribution}

\subsection{Experiment Design}

\subsubsection{Research Questions}
To explore the impact of malware family types and composition on malware detection in comprehensive, we further divide RQ2 into the following sub-questions.
\begin{itemize}
    \item[RQ2.1] Does the number of malware families in a given dataset affect the detection results? 
    \item[RQ2.2] Assuming the number of malware families in a given dataset is stable, does the composition of malware families affect the malware detection results?
    \item[RQ2.3] 
    Considering the uneven distribution of malware families in the existing dataset (see Table~\ref{table:existingdataset}), does the emphasis on certain families affect the overall detection results?
\end{itemize}

\begin{figure*}[h]
  \centering
  {
\includegraphics[width=0.32\linewidth]{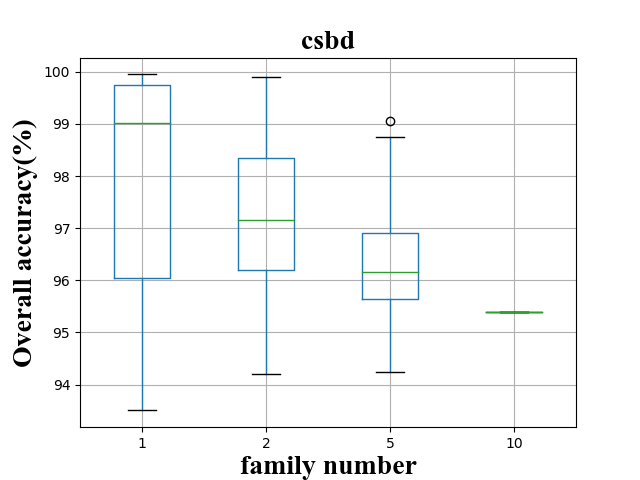}
\includegraphics[width=0.32\linewidth]{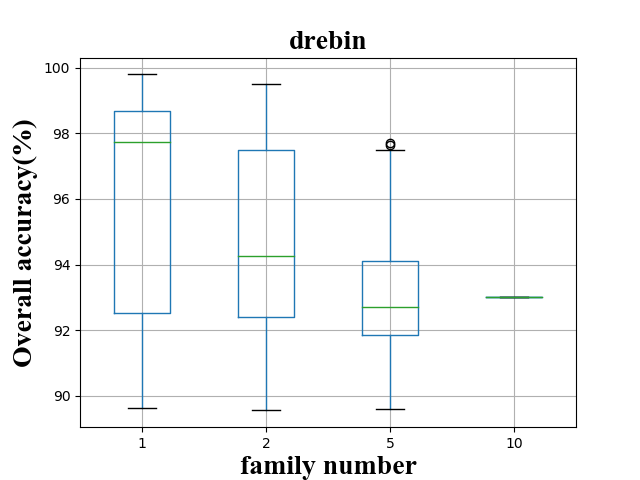}
\includegraphics[width=0.32\linewidth]{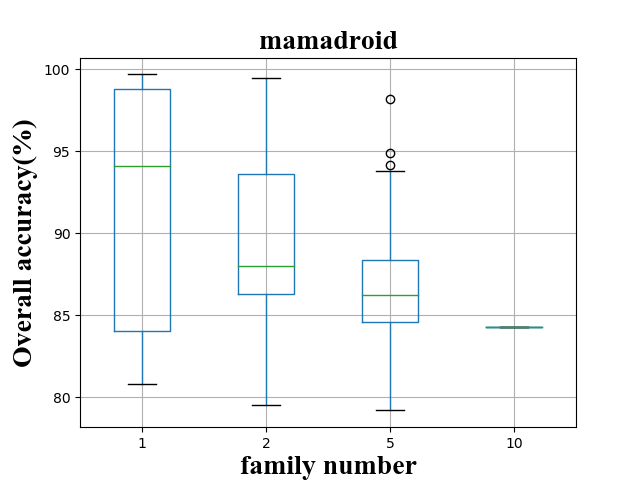}
}
  \caption{The malware detection results for different number of malware families in a given dataset.}
  \label{fig:RQ2.1}
\end{figure*}

\subsubsection{Target Malware Families}
We randomly select 10 popular malware families from the malware corpus mentioned in Section~\ref{sec:design} to investigate the impact of the types and compositions of malware families on malware detection. Note that, we did not consider adware related malware families in this study. 
Table \ref{tab:famname} shows the family name of the selected malware and its proportion in our malware corpus. These malware families have diverse malicious behaviors and different levels of prevalence in the wild. 
Note that in the sets of experiments in RQ2, all the selected malware samples were tagged by at least five engines on VirusTotal to eliminate the potential bias introduced by the labelling thresholds (see Section~\ref{sec:rq1}). 


\begin{table}[t]
\centering
\caption{10 Families Selected in RQ2}
\label{tab:famname}
\begin{tabular}{c|c|c|c}
\hline
Family Name & Percentage\% & Family Name & Percentage\% \\ \hline
artemis     & 8.48\%       & fakeapp     & 0.60\%       \\
generisk    & 3.23\%       & scamapp     & 0.27\%       \\
genbl       & 2.34\%       & mobeleader  & 0.26\%       \\
hifrm       & 1.71\%       & hamob       & 0.22\%       \\
plankton    & 0.84\%       & stopsms     & 0.22\%       \\ \hline
\end{tabular}%
\end{table}

\subsubsection{Dataset Construction}
To answer RQ2.1 and RQ2.2, we have curated a number of malware datasets with different numbers of malware families. 
Each malware family is evenly distributed in the dataset. 
The malware datasets include only one family (10 datasets of 10 selected malware families), two families (select \emph{2} families from the 10 families, and perform \emph{45} ($\binom{2}{10}$) groups of experiments), five families (select \emph{5} families from the 10 families, and perform \emph{252} ($\binom{5}{10}$) groups of experiments) and ten families (all 10 selected families).
Based on these 308 different malware datasets, we explore the impact of the malware family distribution on malware detection.

To investigate the uneven distribution of malware families raised in RQ2.3, we further construct malware datasets that are dominated by some malware families.
To be specific, for the selected 10 families, we randomly select two families (\emph{45} ($\binom{2}{10}$) groups of experiments) and regard them as the two major families, i.e., the two major families occupy 60\% of the malware samples in the dataset (each occupies 30\% of the malware samples), and the remaining eight families occupy 40\% of the dataset (each family occupies 5\% of the dataset). We explore whether the imbalance distribution affects the results of malware detection based on these datasets.






\subsection{RQ2.1 Does the number of malware families matter?}

Figure~\ref{fig:RQ2.1} shows the distribution of the accuracy of the experiment results for the different number of malware families in the dataset. With the number of the malware families growing, the average detection accuracy of each malware detection method decreases. On average, malware detection works best in the case when the dataset contains only one malware family. The trends of the three approaches are similar. The difference between accuracy caused by the number of malware families in the dataset is 2.61\%, 3.01\%, and 7.49\% respectively for the three algorithms.


\begin{tcolorbox}[title= \textbf{Brief Summary}, left=2pt, right=2pt, top=2pt,bottom=2pt]
On the condition of the number of samples in a malware dataset is stable, the number of families matters. More families, more diversity. In general, with the increasing of the number of malware families, the accuracy of machine-learning based classifiers would decrease. Different algorithms have different sensitivities to the number of malware families in a dataset.
\end{tcolorbox}

\begin{figure}[h]
  \centering
  \includegraphics[width=0.99\linewidth]{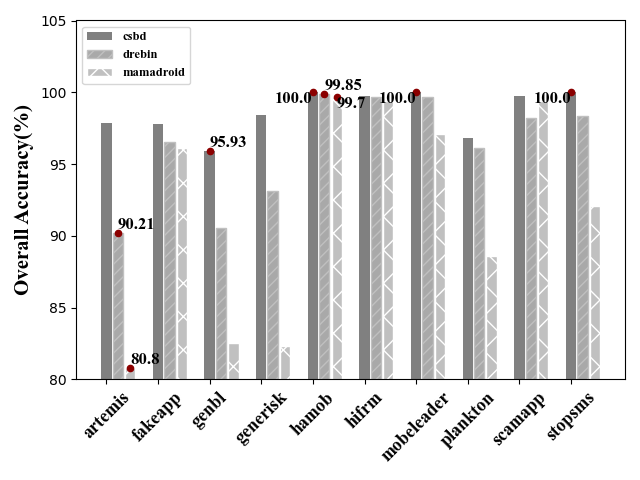}
  \vspace{-0.15in}
  \caption{The accuracy of malware detection on different families.}
  \vspace{-0.15in}
  \label{fig:rq2.2all}
\end{figure}

\subsection{RQ2.2 Does the composition of malware families matter?}

RQ2.1 and RQ2.2 use the same dataset but do not focus on the same issues. RQ2.1 focuses on the impact of the number of malware in the dataset, while RQ2.2 pays more attention on the composition of malware families in the dataset.
We consider the case that the number of families in the malware dataset is 1, 2, and 5. We further investigate the differences in their performances based on our constructed datasets.

\begin{figure*}[h]
  \centering
  {
\includegraphics[width=0.3\linewidth]{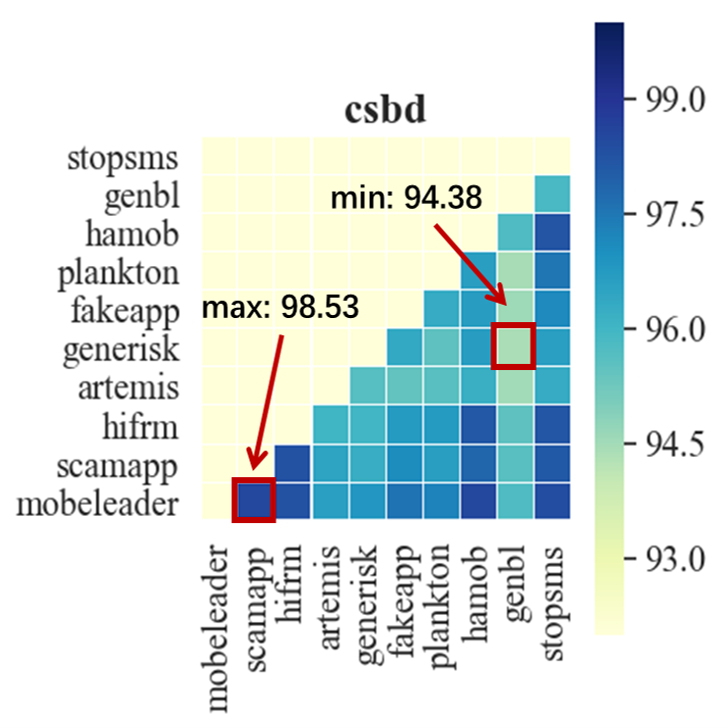}
\includegraphics[width=0.3\linewidth]{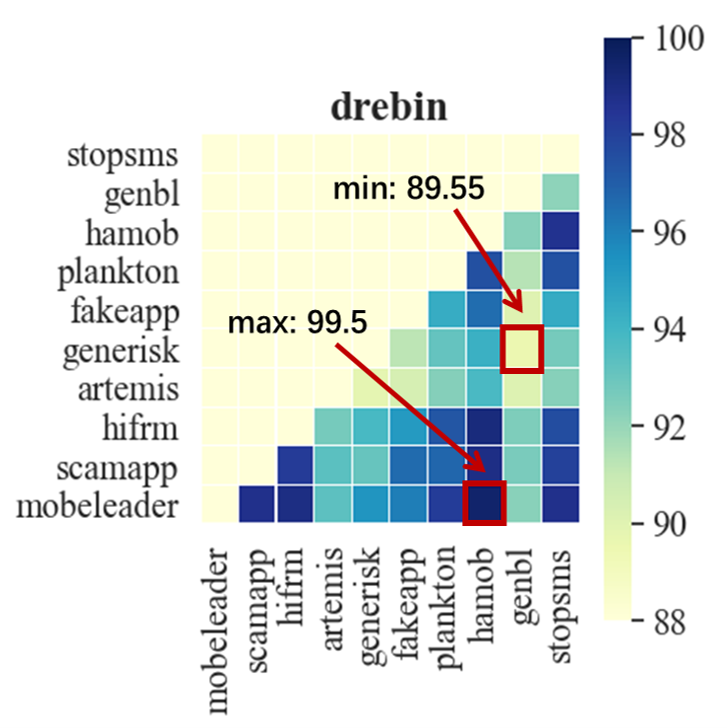}
\includegraphics[width=0.3\linewidth]{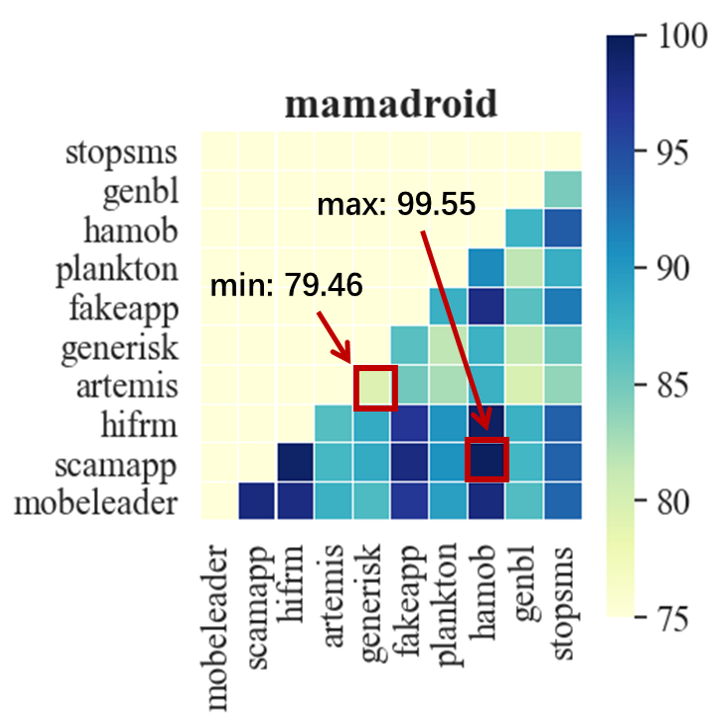}
}
  \caption{Experiment results on malware dataset composed of two families. Each cell represents the overall accuracy of the malware detection approach on the dataset composed of the corresponding two families (x-axis and y-axis).}
  \label{fig:rq2.2}
\end{figure*}

\subsubsection{Single Family}
As shown in Figure~\ref{fig:rq2.2all}, for the constructed 10 malware datasets (with each having one single-family), the detection accuracy of the trained classifiers has noticeable differences. 
The gap of the malware detection accuracy could be up to 4.07\%, 9.64\%, and 18.9\% respectively for \texttt{csbd}, \texttt{drebin}, and \texttt{mamadroid}, respectively. 
We further perform inter-approach analysis.
All these three detection approaches have their ``favorable'' families.
For example, as shown in Figure~\ref{fig:rq2.2all}, it is apparently that \texttt{drebin} performs better than \texttt{mamadroid} in most cases. Nevertheless, the malware detection accuracy of \texttt{mamadroid} is higher than \texttt{drebin} on the \texttt{scamapp} dataset. Thus, for a malware dataset dominated by a single family, the evaluation results might be quite bias.

\subsubsection{Two Families}
We next investigate the malware dataset composed of two families. Figure \ref{fig:rq2.2} shows the overall results.
Deeper color represents higher accuracy. 
We can observe that different combinations of families result in huge differences in accuracy under the same malware detection method. 
The gap of the malware detection accuracy could be over 20\% even for the same malware detection approach (4.15\%, 9.95\%, and 20.09\% respectively between the highest and the lowest for \texttt{csbd}, \texttt{drebin}, and \texttt{mamadroid}). 
We further notice that all of these three approaches can achieve the best results under different combinations. 
For the classifiers trained using \texttt{csbd}, the one that relies on the dataset contains \texttt{scamapp} and \texttt{mobeleader} families performs the best. As for \texttt{drebin}, the dataset containing \texttt{mobeleader} and \texttt{hamob} families performs the best. The dataset containing \texttt{hamob} and \texttt{scamapp} families performs the best for \texttt{mamadroid}. 
It further indicates that the evaluation of the malware detection approach is highly relevant to the constructed dataset.

\begin{figure}[h]
  \centering
  \includegraphics[width=0.99\linewidth]{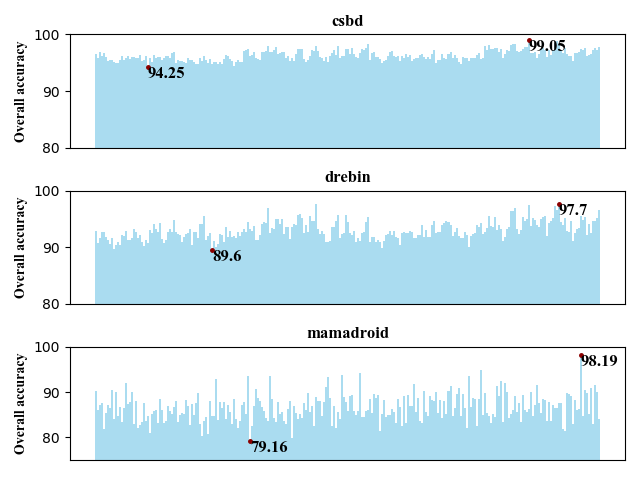}
  \caption{Experiment results on malware dataset composed of five families. Note that we have created 225 groups of datasets in total.}
  \label{fig:rq2.2all-5}
\end{figure}

\begin{table*}[t]
\centering
\caption{5 sets of families with the highest/lowest accurary for mamadroid}
\label{tab:5-mamadroid}
\resizebox{\textwidth}{!}{%
\begin{tabular}{c|c|cc}
\hline
\multicolumn{2}{c|}{Highest Five}                            & \multicolumn{2}{c}{Lowest Five}                                                  \\ \hline
Accuracy\% & Families                                       & \multicolumn{1}{c|}{Accuracy\%} & Families                                      \\ \hline
98.19\%    & fakeapp,hifrm,mobeleader,scamapp,hamob  & \multicolumn{1}{c|}{79.15\%}     & plankton,generisk,fakeapp,artemis,genbl \\
94.87\%     & hifrm,mobeleader,scamapp,hamob,stopsms & \multicolumn{1}{c|}{79.95\%}     & plankton,generisk,artemis,genbl,hamob      \\
94.22\%    & fakeapp,hifrm,scamapp,hamob,stopsms  & \multicolumn{1}{c|}{80.31\%}    & plankton,generisk,artemis,genbl,stopsms \\
93.82\%     & fakeapp,hifrm,mobeleader,hamob,stopsms    & \multicolumn{1}{c|}{80.71\%}   & plankton,generisk,artemis,genbl,scamapp \\
93.57\%    & plankton,fakeapp,mobeleader,scamapp,hamob      & \multicolumn{1}{c|}{80.91\%}    & plankton,generisk,hifrm,artemis,genbl    \\ \hline
\end{tabular}%
}
\end{table*}

\begin{figure*}[h]
  \centering
  {
\includegraphics[width=0.3\linewidth]{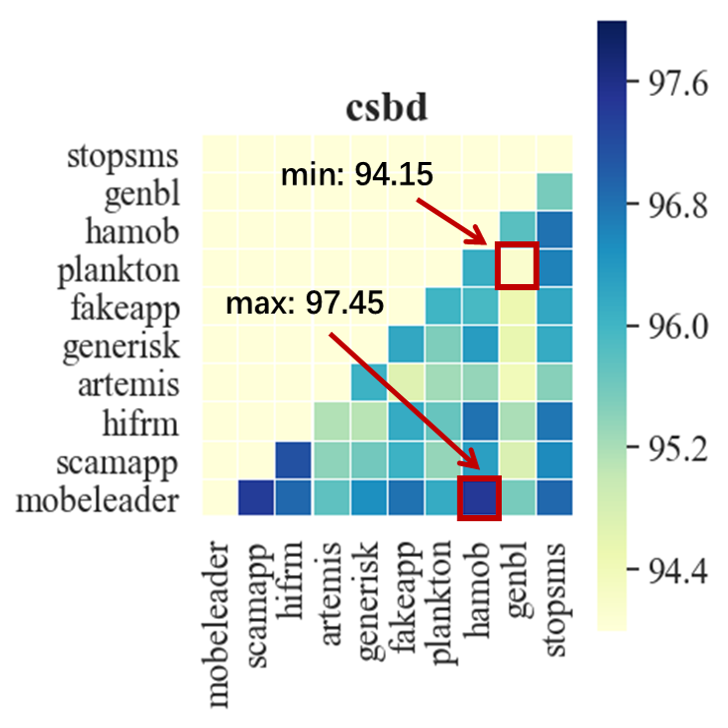}
\includegraphics[width=0.3\linewidth]{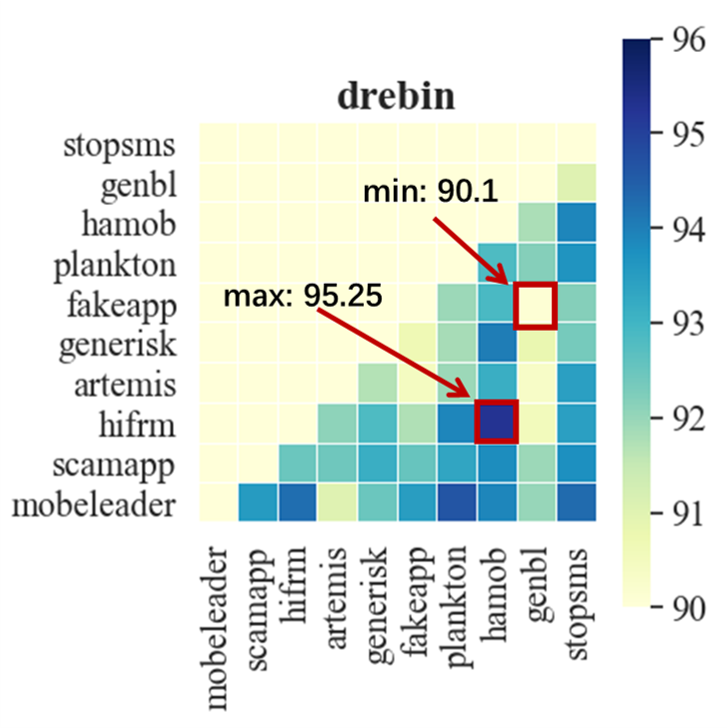}
\includegraphics[width=0.3\linewidth]{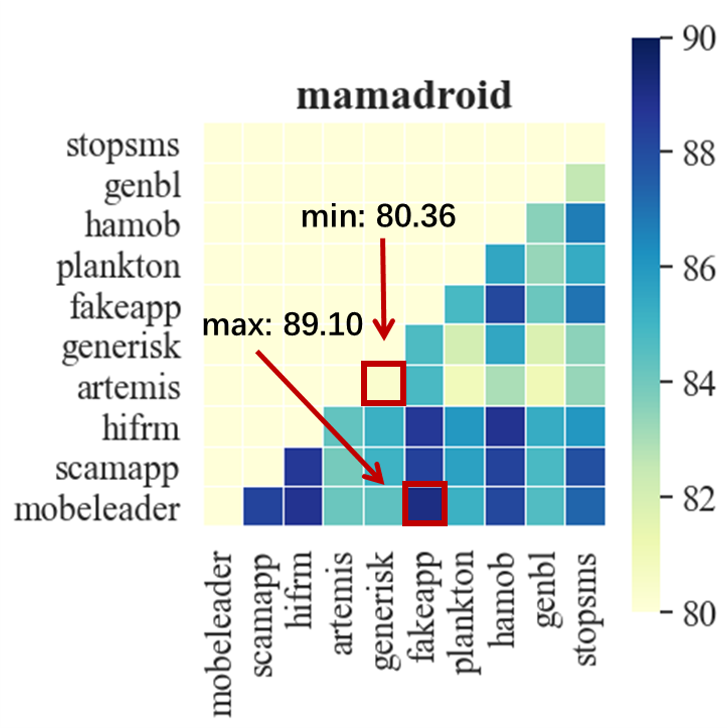}
}
\caption{The impact of the imbalance of malware families.}
\label{fig:64-accuracy}
\end{figure*}

\subsubsection{Five Families}
We next apply the malware detection approaches to datasets which are composed of five malware families, with 252 groups of malware benchmarks in total.
Figure \ref{fig:rq2.2all-5} shows the overall distribution of the results. It is surprising to see that, for a single approach, the difference between the best and the worst results could achieve 4.8\% for \texttt{csbd}, 8.1\% for \texttt{drebin}, and 19.04\% for \texttt{mamadroid} (the best and the worst results are highlighted on Figure~\ref{fig:rq2.2all-5}). 
For example, Table \ref{tab:5-mamadroid} shows the top-5 classifiers with the highest and the lowest accuracy for classifiers trained using \texttt{mamadroid}. 
The malware detection performance of \texttt{csbd} is relatively stable, and the variance of accuracy is 0.82\%. The variances of \texttt{drebin} and \texttt{mamadroid} are 2.58\% and 8.92\%, respectively.

We further perform inter-approach analysis.
\texttt{Csbd} always performs the best under the 252 groups of malware benchmarks.  
In most cases, the accuracy of \texttt{drebin} is higher than that of \texttt{mamadroid}. However, although the accuracy of \texttt{mamadroid} is the most unstable, \texttt{mamadroid} could also perform better than \texttt{drebin} in some cases. For example, on the dataset of the composition of \texttt{generisk}, \texttt{hifrm}, \texttt{artemis}, \texttt{mobeleader} and \texttt{genbl}, \texttt{drebin} achieves 8.25\% of  accuracy higher than \texttt{mamadroid} (82.87\% VS. 91.12\%), while on the dataset of the composition of \texttt{fakeapp}, \texttt{hifrm}, \texttt{mobeleader}, \texttt{scamapp} and \texttt{hamob}, \texttt{mamadroid} achieves 2.69\% of  accuracy higher than \texttt{drebin} (98.19\% VS. 95.5\%)

\begin{tcolorbox}[title= \textbf{Brief Summary}, left=2pt, right=2pt, top=2pt,bottom=2pt]
Our findings highlight that, in the research of malware detection, it is easy to ignore the dataset bias of family composition, which leads to an excessive interpretation of malware detection approaches.
The composition of malware families can have great influences on the malware detection performance when the number of malware families in a given dataset is stable. The superiority of malware detection approach is different under various combinations of malware families. One can even manipulate the composition of the dataset to make one specific approach achieve the best results. 
\end{tcolorbox}



\subsection{RQ2.3 Does the imbalance of malware families matter?}

As aforementioned, we construct malware datasets that are dominated by two random malware families and perform 45 groups of experiments. The results are presented in Figure~\ref{fig:64-accuracy}.

\subsubsection{Intra-approach Analysis}
As each dataset has two dominating families, each cell in the Matrix indicates the corresponding major families. For the considered three approaches, the difference between the highest and the lowest accuracy is 3.3\% (\texttt{csbd}), 5.15\% (\texttt{drebin}) and 8.73\% (\texttt{mamadroid}), respectively. 

\begin{figure*}[t]
\centering
\includegraphics[width=0.3\linewidth]{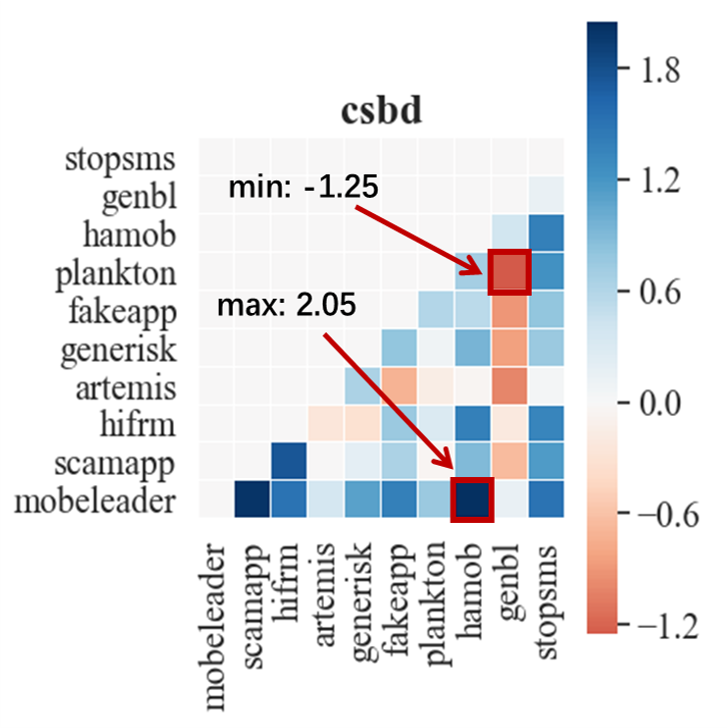}
\includegraphics[width=0.3\linewidth]{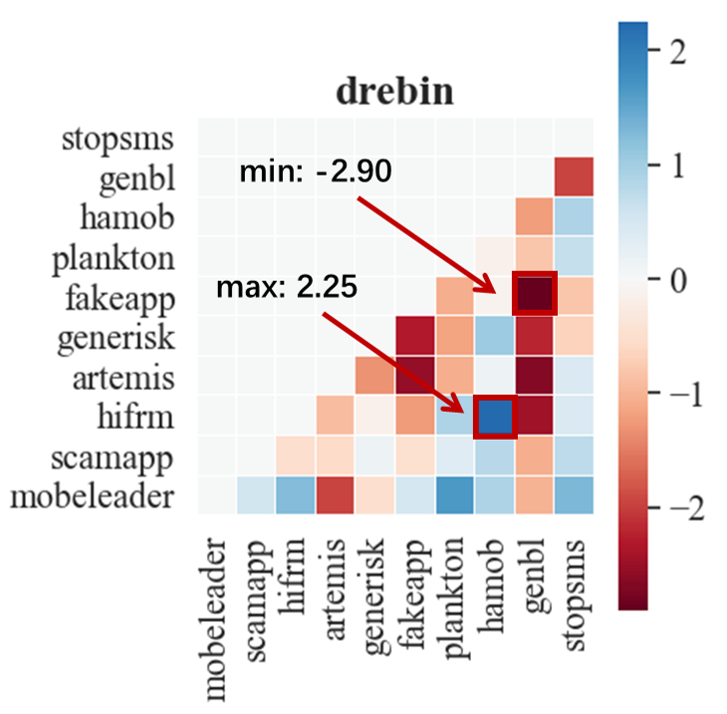}
\includegraphics[width=0.3\linewidth]{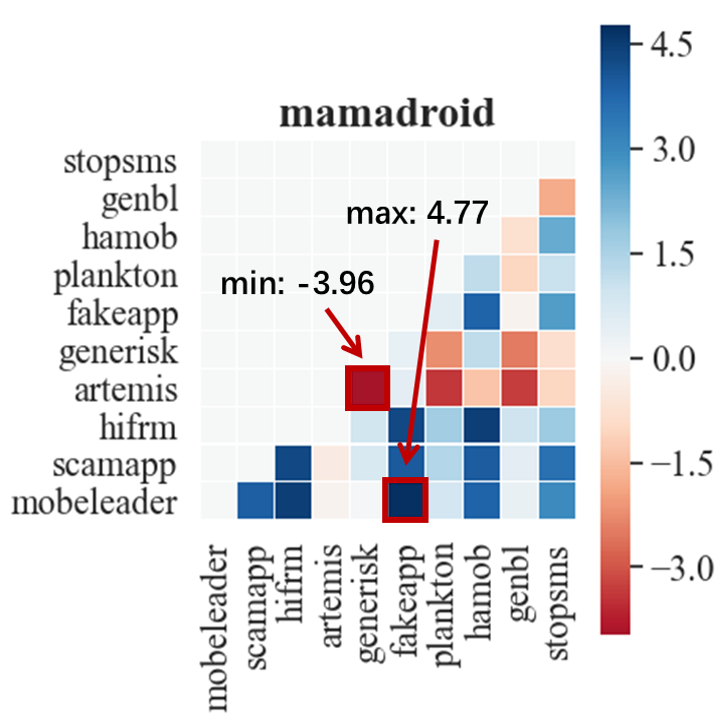}
\caption{The accuracy difference with between imbalance distribution datasets and average distribution datasets.}
\label{fig:difference}
\end{figure*}

For \texttt{csbd}, as shown in Figure \ref{fig:difference}, 32 of 45 combinations are higher than the average distribution datasets. 
The minimum accuracy difference is -1.25\%, and the maximum is 2.05\%. 
For \texttt{drebin}, 18 of them are higher than the average distribution datasets, and the minimum accuracy difference is -2.9\%, and the maximum is 2.25\%. For \texttt{mamadroid}, 30 of them are higher than the average distribution datasets, and the minimum accuracy difference is -3.96\%, and the maximum is 4.77\%. 

\subsubsection{Inter-approach Analysis}
In general, \texttt{csbd} performs the best, follewed by \texttt{drebin} and then \texttt{mamadroid}. Further, the performance gaps among these approaches are highly related to the dominating families. For example, when the two dominating families of the dataset are \texttt{hifrm} and \texttt{hamob}, \texttt{drebin} achieves 1.9\% of accuracy higher than \texttt{mamadroid} (90.7\% VS. 88.8\%), while \texttt{drebin} could achieve 13.3\% of accuracy higher than \texttt{mamadroid} (93.7\% VS. 80.4\%) when the two dominating families of the dataset are \texttt{generisk} and \texttt{artemis}.


\begin{tcolorbox}[title= \textbf{Brief Summary}, left=2pt, right=2pt, top=2pt,bottom=2pt]
The emphasis on certain families can affect malware detection performance. The impact on the same malware detection differs when the malware families with a large share in the dataset are different. Also, the effect of increasing or decreasing the percentage of some certain malware families in the dataset does not have the same effect with different algorithms. Therefore, the imbalance of malware families can be an evaluation bias of malware detection.
\end{tcolorbox}

\begin{figure}[t]
\centering
\includegraphics[width=0.99\linewidth]{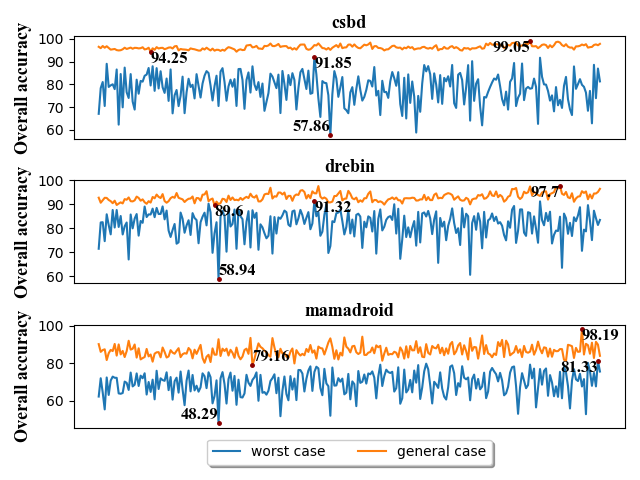}
\caption{The impact of sample partition methods.}
\label{fig:difference2}
\end{figure}


\section{The Impact of Sample Partition}
\label{sec:rq4}
\subsection{Experimental Design}
We further design experiments to evaluate the impact of the sample partition methods in cross-validation. We consider the ideal case and the extreme case of the sample partition methods in cross-validation to present the comparisons more visually.
In the ideal case, each malware family in the dataset is evenly distributed, and the types and numbers of malware families in the training and validation sets are identical. It can ensure that the data distribution of the malware detection methods in the training and testing set is completely consistent.
As comparison, one extreme case is that the verification set does not even contain the malware families in the training set. 
We compare the ideal case with the extreme case using 5-Fold cross validation. We re-use the same 252 groups of malware benchmarks from RQ2.2. For the extreme case, the malware dataset is partitioned into 5 sub-sets, and each sub-set contains a single family, which means it does not have malware family overlap with other sub-sets. Figure~\ref{fig:difference2} shows the overall results for the three approaches.

\subsection{Analysis}
\textbf{Intra-approach Analysis}
If the sample partition methods in cross validation is not randomly partitioned, all the three methods could not ensure the malware detection performance. 
The malware detection methods could not be fair evaluated if the bias of sample partition exists. Figure \ref{fig:difference2} shows the overall results. As expected, for all the three malware detection approaches, their performance decreased greatly on the extreme case. The gap between the worst case and the ideal case could be 37.29\%, 31.16\%, and 44.58\% for \texttt{csbd}, \texttt{drebin}, and \texttt{mamadroid} respectively. 

\textbf{Inter-approach Analysis}
The sample partition method has a huge influence on the comparison of different malware detection approaches. In the ideal case, \texttt{csbd} always performs the best when comparing the accuracy of the three algorithms. In the worst case, it is hard to tell which malware detection approach performs the best. It is found that \texttt{drebin} performs the best 179 times and \texttt{csbd} performs the best 73 times in the worst case.



\begin{tcolorbox}[title= \textbf{Brief Summary}, left=2pt, right=2pt, top=2pt,bottom=2pt]
Our exploration suggests that even using the same dataset, data partition methods used in the cross-validation could introduce substantial impact to the overall results. One can even manipulate the training/testing data partition method to make one specific approach perform the best. 
We argue that there might be research bias when evaluating their approaches. 
\end{tcolorbox}

\section{Discussion}
\label{Discussion}

\subsection{Implication}
Our observations in this paper can provide practical implications. First, we observe that the effectiveness of malware detection approaches is highly related to the dataset used in the evaluation and the experiment results are inflated due to research bias introduced by the variability of the malware dataset and the methods to use the dataset.
The results can be inconsistent and even misleading, i.e., the superiority of malware detection methods under different criteria is different. In other words, evaluating whether a method is good enough is closely relevant to the chosen dataset. This, however, has never been mentioned and carefully explained in the previous thousands of malware detection papers. 
We admit that different malware detection approach may have their own advantages and their usage scenarios, however,
we argue that, a machine learning fairness framework for malware detection is needed to highlight the pros and cons of each approach. Providing reasonable, meaningful and explainable results is better than only reporting a high detection accuracy with vague dataset and experimental settings.

\subsection{Limitation}
Our study carries several limitations. First, we have investigated three aspects of the dataset bias in Android malware detection. Although it is still possible that the family distribution is unlimited and other situations of dataset bias exist. Nevertheless, we believe our design has covered most of the situations of dataset bias. Second, we randomly select 10 families to do experiments in this paper. However, various malware families exist in the industry environment. We believe that similar conclusions can be obtained by using other families. Finally, due to the resource limitations, our research is mainly based on three malware detection methods. We make efforts to investigate the differences between the three methods to form a general law of the dataset bias in Android malware detection.

\section{Conclusion}
\label{Conclusion}
In this paper, we make the first systematic study of the dataset bias in Android malware detection. Specifically, we focus on the variability of the malware dataset and the methods to use the dataset. We make effort to curate a number of malware benchmarks under different conditions and evaluate their performance gap of malware detection using existing techniques. The experiment results suggest that, using different criteria to create and use the malware dataset can have a misleading impact on evaluations. We argue that  a machine learning fairness framework for malware detection is needed to highlight the pros and cons of each approach. Providing reasonable, meaningful and explainable results is better than only reporting a high detection accuracy with vague dataset and experimental settings. 


\bibliographystyle{IEEEtran}
\bibliography{main}

\vspace{12pt}

\end{document}